\documentclass[a4paper]{article}

\usepackage[margin=1in]{geometry} 

\usepackage{amsmath}
\usepackage{amsthm}
\usepackage{amssymb}

\usepackage[utf8]{inputenc}
\usepackage{hyperref}
\hypersetup{
	unicode,
	pdfauthor={},
	pdftitle={Programmable Integrated Photonics for Topological Hamiltonians},
	pdfsubject={},
	pdfkeywords={},
	pdfproducer={LaTeX},
	pdfcreator={pdflatex}
}

\theoremstyle{plain}

\theoremstyle{definition}

\usepackage{graphicx, color}
\graphicspath{{fig/}}

\usepackage{algorithm, algpseudocode} 
\usepackage{mathrsfs} 

\usepackage{lipsum}
\usepackage{cite}

\newcommand{\rem}[1]{}
\newcommand{\add}{\textcolor{black}}

\title{Programmable Integrated Photonics for Topological Hamiltonians}
\author{Mehmet Berkay On$^{1,2}$, Farshid Ashtiani$^{1}$, David Sanchez-Jacome$^{3}$, Daniel Perez-Lopez$^{3}$, \and S. J. Ben Yoo$^{2}$, and Andrea Blanco-Redondo$^{1,4,*}$}

\date{
	$^1$\textit{Nokia Bell Labs, 600 Mountain Ave, New Providence, NJ 07974, USA} \\%
	$^2$\textit{University of California Davis, Department of Electrical and Computer Engineering, One Shields Avenue, Davis, CA 95616, USA}\\%
    $^3$\textit{iPronics Programmable Photonics, Avenida Blasco Ibanez 25, Valencia 46010, Spain}\\%
    $^4$\textit{CREOL, The College of Optics and Photonics, University of Central Florida, Orlando, FL 32816, USA}\\%
    $^*$andrea.blancoredondo@ucf.edu\\[2ex]%
}

\begin{document}
	\maketitle

	\begin{abstract}
    A variety of topological Hamiltonians have been demonstrated in photonic platforms, leading to fundamental discoveries and enhanced robustness in applications such as lasing, sensing, and quantum technologies. To date, each topological photonic platform implements a specific type of Hamiltonian with inexistent or limited reconfigurability.  Here, we propose and demonstrate different topological models by using the same reprogrammable integrated photonics platform, consisting of a hexagonal mesh of silicon Mach-Zehnder interferometers with phase-shifters. We specifically demonstrate a one-dimensional Su-Schrieffer-Heeger Hamiltonian supporting a localized topological edge mode and a higher-order topological insulator based on a two-dimensional breathing Kagome Hamiltonian with three corner states.  These results highlight a nearly universal platform for topological models that may fast-track research progress toward applications of topological photonics and other coupled systems.
	\end{abstract}

	
	\section{Introduction}
The field of topological photonics \cite{RevModPhys.91.015006,Price2022} has gained tremendous traction in the last 15 years thanks to its unraveling of novel fundamental phenomena in topological physics as well as its potential to deliver robustness against certain types of defects and disorder for integrated photonic devices \cite{shalaev2019robust,arora2021direct} such as lasers \cite{St-Jean2017,bahari2017nonreciprocal, Bandres2018,Contractor2022} and quantum information platforms \cite{Mittal2018,Blanco-Redondo2018,Wang2019,Mittal2021,Doyle2022}.  The origins of topological photonics stem from the discovery of topological insulators in condensed matter physics \cite{Klitzing1980, Kivelson1982}, where materials that are insulating in their bulk can conduct electricity without dissipation on their edges. These concepts were translated into photonics platforms \cite{Haldane2008,wang2009observation}, where topology refers to a quantized property that describes the global behavior of the wavefunctions in a dispersion band.
A key feature of topological photonics is the existence of modes that live on the edge of photonic materials with different topologies and that show resilience to certain types of disorder. These edge modes have been demonstrated in a variety of platforms, from one-dimensional (1D) arrays of waveguides \cite{Malkova2009,Zeuner2015,Blanco-Redondo2016} or resonators \cite{Poli2015} with chiral symmetries, to two dimensional (2D) lattices of helical waveguides \cite{rechtsman2013photonic} and ring resonators with asymmetric couplings \cite{Hafezi2013}, all the way through bianisotropic metamaterials \cite{Khanikaev2012} and quasicrystals \cite{Verbin2013}.
While the majority of topological photonics platforms presented to date have a static character, a number of reconfigurable topological photonic insulators have been experimentally realized in the last few years \cite{Cheng2016,Zhao2019, Cao2019}, as well as analogous concepts in acoustics \cite{Xia2018, Darabi2020} and plasmonics \cite{You2021}. However, the reconfigurability in these platforms is limited to rerouting the pathways followed by the guided waves or switching these pathways on and off, while the type of Hamiltonian implemented in a given physical platform is fixed.


In parallel, programmable integrated photonic platforms have enabled fast development of a wide range of circuit architectures through real-time reconfiguration of a general-purpose photonic circuit via software programming \cite{Bogaerts2020}. Such systems typically consist of a 2D mesh of silicon photonics Mach-Zehnder interferometers (MZIs) whose transfer matrix can be programmed by adjusting the embedded phase shifters. This enables the reconfiguration of light paths through the mesh and the implementation of linear optical operations by interfering signals from different paths \cite{Perez-Lopez2020}, showing a ground-breaking potential for communications, machine learning \cite{Harris:18} and quantum information processing \cite{Clements:16} among other applications.


Here, we propose and experimentally demonstrate that topological physics can be observed in programmable integrated photonics platforms. Importantly, virtually any topological model can be implemented in programmable integrated photonic platforms that allow for exquisite reconfigurable control of the hopping strength and hopping phase between elements, as well as of the real and imaginary part of the onsite energies. 
To illustrate this, we use a commercial programmable platform (\emph{iPronics' Smartlight Processor}) to show robust localization of edge modes in a dimer chain of resonators resembling the Su-Schrieffer-Heeger (SSH) model \cite{Su1979} and of higher-order topological modes (corner modes) in a 2D breathing Kagome lattice \cite{Ezawa2018} of resonators. Reprogrammable silicon photonic meshes represent a nearly universal test-bed for topological photonics, including non-Hermitian topological photonics \cite{guo2009observation,ruter2010observation,ozdemir2019parity}, that could greatly accelerate fundamental discoveries as well as the development of applications.



\section{Integrated Programmable Mesh}

\begin{figure}[ht!]
\centering\includegraphics[width=\linewidth]{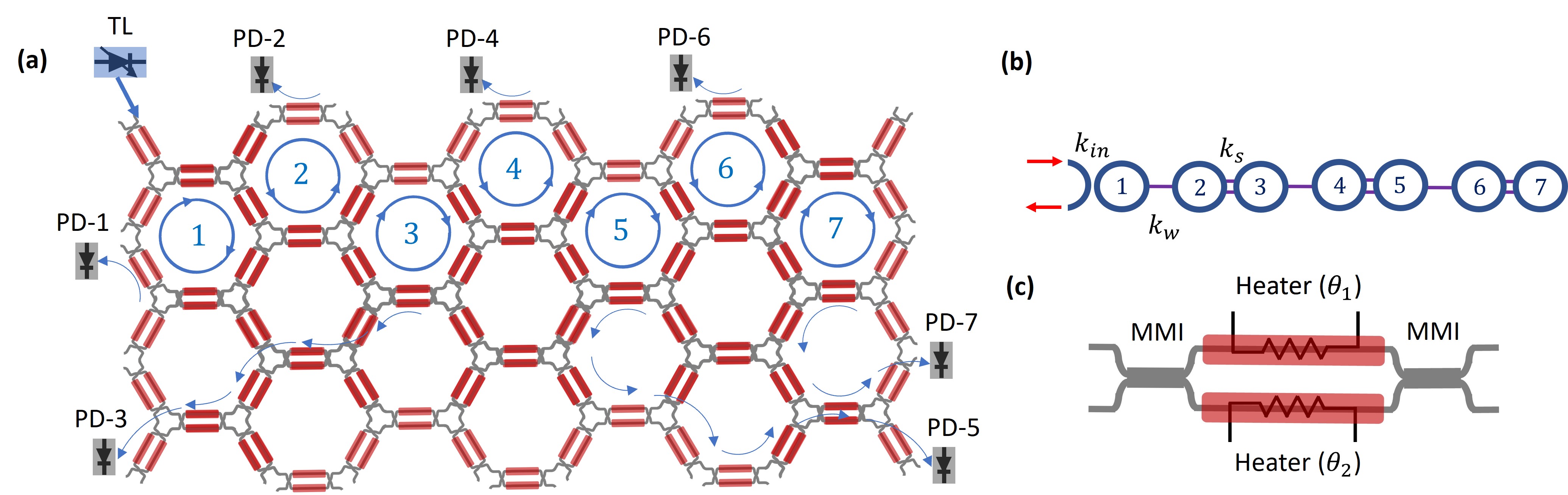}
\caption{(a)Schematic of the programmable mesh on iPronics' Smartlight Processor and reconfiguration for 7 coupled ring resonators; TL: off-chip tunable laser, PD: off-chip photodetector (b) Implemented 1D SSH model. (c) Programmable unit cell in detail.}
\label{fig:hw_mesh}
\end{figure}

A schematic view of the programmable silicon photonics chip used in our experiments is shown in Fig. \ref{fig:hw_mesh} (a). It consists of a hexagonal mesh of programmable unit cells (PUCs), where each PUC is formed by a 2x2 Mach Zehnder interferometer (MZI) with a thermo-optic phase shifter in each arm, as depicted in Fig. \ref{fig:hw_mesh}(c) \cite{Perez:16}. The two optical inputs enter a 50/50 multimode interference (MMI) coupler followed by two thermo-optic phase shifters to adjust the optical phase shift of each arm. Another 50/50 MMI coupler combines the two phase adjusted signals and provides the PUC outputs. By controlling the phases imparted on each arm $\theta_1$ and $\theta_2$ one can realize any 2x2 complex unitary transfer matrix 

\begin{equation}
T(\theta_1,\theta_2)=e^{j\phi}\begin{bmatrix} \cos(\Delta) & -\sin(\Delta)\\\sin(\Delta) & \cos(\Delta)\end{bmatrix}
\end{equation}
with

\begin{equation}
\phi=\frac {\theta_1+\theta_2}{2}
\end{equation}
representing a common phase to the two output signals and 

\begin{equation}
\Delta=\frac {\theta_1-\theta_2}{2}
\end{equation}
determining the power splitting ratio between signals. Therefore, by programming the phase settings of the mesh PUCs, the optical signal can be routed into desired paths and arbitrary photonic circuit configurations can be realized.

To approach the realization of topological physics in programmable meshes we reconfigure the programmable cells of the mesh to create lattices of ring resonators with carefully engineered resonant frequencies and coupling rate between them. Note that, thanks to its interconnection profile, the hexagonal mesh allows for the programming of optical cavities and better resolution when compared to alternative lattice mesh designs \cite{Perez:16,lopez2019programmable}, and it is, therefore, better suited to implement topological Hamiltonians. The smallest possible ring resonator in this hexagonal mesh consists of six PUCs, as schematically depicted by the blue circumferences in Fig. \ref{fig:hw_mesh}(a). The PUCs shared between adjacent rings are programmed to determine the coupling rate (and if desirable the coupling phase) between the two rings. The power in each ring can be monitored by tapping a small amount of the power out of the ring to a monitoring photodiode, as depicted by the blue arrows exiting the lattice. 


In particular, we have chosen to implement two different models to demonstrate the potential of programmable photonics to explore topological physics: a 1D SSH model and a 2D breathing Kagome lattice. Due to the size of the currently available hardware mesh, a rectangular arrangement of 72 PUCs shown in Fig.\ref{fig:hw_mesh}(a), only the 1D SSH model could be experimentally tested in the hardware. Nonetheless, we have implemented the 2D Kagome in a realistic simulator \cite{Perez-Lopez2022} of the mesh and we highlight that the size and shape of the lattice is well within the scalability scope of current technology.




\section{1D Topological Photonics in the Programmable Mesh}
We start by implementing the simplest topological model, the dimer chain, also referred to as the SSH model \cite{Su1979}, which relies on an alternate pattern of weak and strong coupling between sites and was demonstrated in optical experiments in 2009 in an optically-induced superlattice \cite{Malkova2009}. Since then, many optical implementations of the SSH have been proposed: from femtosecond laser written waveguides in glass \cite{Zeuner2015} to silicon photonics waveguides \cite{Blanco-Redondo2016}, all the way to microwave resonators \cite{Poli2015} and others. All of these demonstrations have shown little to none reconfigurability. Here, we implement the SSH model in a programmable mesh by arranging the mesh into a bipartite lattice of seven ring resonators, as schematically depicted in Fig. \ref{fig:hw_mesh}(b). The experimental realization of this model on the silicon photonics programmable platform is marked by blue circumferences in Fig. \ref{fig:hw_mesh}(a). 

The Hamiltonian describing this system of seven rings is given by
\begin{equation}
H=\left[k_w\sum_{n\in\{1,3,5\}} a_n^\dag a_{n+1} + k_s\sum_{n\in\{2,4,6\}} a_n^\dag a_{n+1}\right]+H.c.
\end{equation}
where $k_w$ and $k_s$ are the strong and weak coupling strengths between sites -- accurately controlled in this experiment by programming the common PUC between rings -- and $a_n^\dag$ and $a_n$ are the creation and annihilation operators on site $n$. 

\begin{figure}[ht!]
\centering\includegraphics[width=\linewidth]{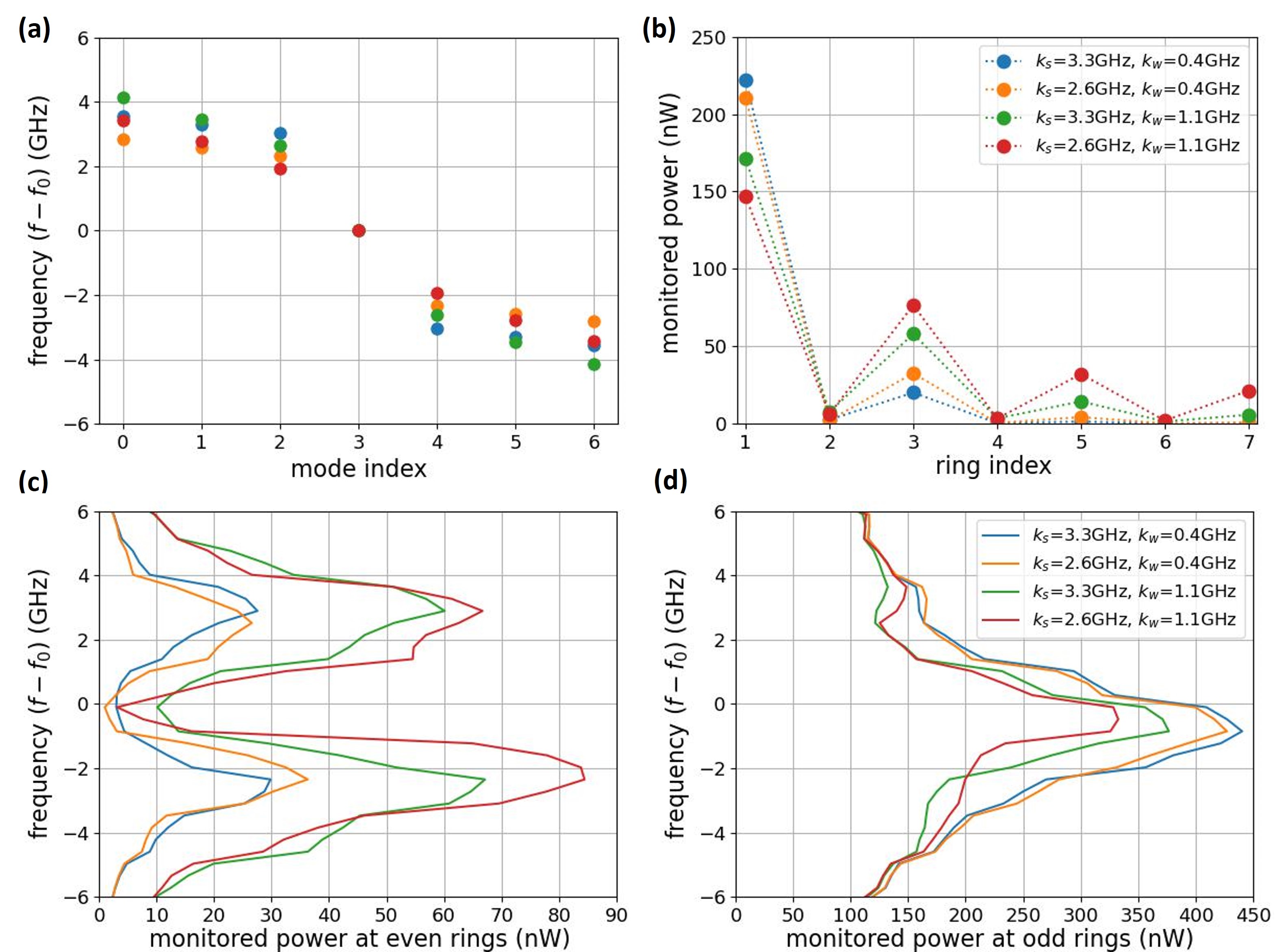}
\caption{(a) Calculated eigenvalues of the Hamiltonian matrices. (b) Monitored powers at the coupled ring resonators from the programmable mesh hardware. (c) Total monitored power spectrum of the even rings of 1D SSH model. (d) Total monitored power spectrum of the odd rings of 1D SSH model.}
\label{fig:1d_result_1}
\end{figure}

The calculated eigenvalues of this lattice, embodied here by the resonant frequencies of the supermodes, are shown in Fig. \ref{fig:1d_result_1} (a) for three different combinations of $k_w$ and $k_s$. The supermodes frequencies are offset to the resonant frequency of the individual ring resonators $f_0=193.396$ THz. Note that slight differences in $f_0$ between rings can be compensated by adjusting the phase shifters in the mesh (see Supplemental Document Section 1.). This lattice is expected to have a bandgap with a topological edge mode localized in ring 1. Stronger dimerization patterns, in other words stronger contrast between $k_w$ and $k_s$, are expected to lead to larger bandgaps and consequently to stronger and more robust localization of the edge mode. Thus, the reprogrammability of the lattice lends us full control over the band gap, the degree of localization and the robustness of the edge mode.

To experimentally prove this, we connect a continuous wave tunable laser to the input port of the mesh and monitor the power in the rings under different conditions. First, we tune the laser wavelength to $f_0$ and monitor the power in each ring using photodiodes. The resulting measurements, shown in Fig. \ref{fig:1d_result_1} (b), exhibit the characteristic modal distribution of the SSH edge modes with a maximum at the edge site and full localization in one of the sublattices, i.e. virtually zero power in the even rings. The measurements also confirm that stronger dimerization patterns lead to stronger localization at the edge. Subsequently, we tuned the input laser frequency within $\pm 6$ GHz around $f_0$ and summed up the power in all the even and odd rings, as shown in Fig. \ref{fig:1d_result_1} (c) and (d), respectively. By looking at the width of the dip around $f_0$ in Fig. \ref{fig:1d_result_1} (c) one can appreciate how the band gap grows with increasing dimerization strength. This is because the only supermode present supported around $f_0$ is the topological edge mode, which is fully localized in the odd rings. Consequently, the peak exhibited around $f_0$ in the odd rings, as shown in Fig. \ref{fig:1d_result_1} (d), correlates strongly with the power in the topological edge mode and it becomes higher with stronger dimerization.

\begin{figure}[ht!]
\centering\includegraphics[width=\linewidth]{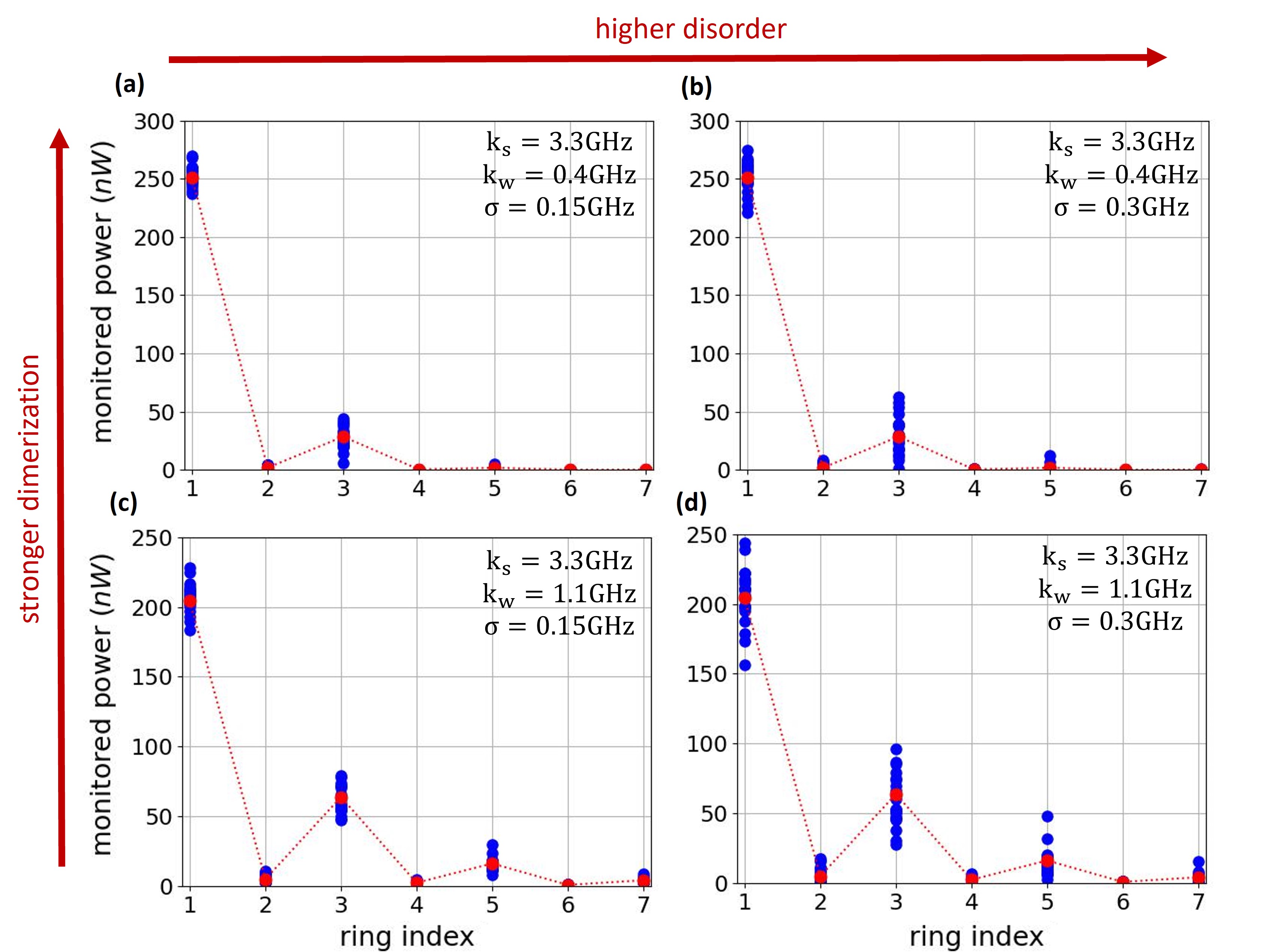}
\caption{Monitored powers at the coupled ring resonators from the programmable mesh hardware with 20 different random perturbation added on the coupling rates for (a) and (b) the \emph{strong dimerization} case and (c) and (d) the \emph{weak dimerization} case.}
\label{fig:1d_result_2}
\end{figure}

Next, we evaluate the robustness of the topological edge state by intentionally introducing perturbations on the coupling strengths. Specifically, twenty random variations are drawn independently from a normal Gaussian distribution around the nominal coupling strength for each pair of rings, $\sim \mathcal{N}(0,\sigma^2)$, where $\sigma$ is the standard deviation. Figure \ref{fig:1d_result_2} shows the power in each ring at $f_0$ for two dimerization patterns -- a \emph{strong dimerization} case with $k_s=3.3$\,GHz, $k_w=0.4$\,GHz in Figs. \ref{fig:1d_result_2} (a) and (b); and a \emph{weak dimerization} case with $k_s=3.3$\,GHz and $k_w=1.1$\,GHz in Figs. \ref{fig:1d_result_2} (c) and (d). For each case, we consider two levels of disorder -- \emph{low disorder} with $\sigma=0.15$\,GHz in Figs. \ref{fig:1d_result_2} (a) and (c) and \emph{high disorder} with $\sigma=0.3$\,GHz in Figs. \ref{fig:1d_result_2} (b) and (d)).  The red dots represent the power in each ring in the absence of deliberately introduced disorder and the blue dots represent the power in the rings when each of the twenty random iterations of disorder is implemented.

We can now quantify the robustness of the topological mode by measuring the standard deviation of the power in the rings under disorder in the coupling. For instance, under \emph{low disorder} (\emph{high disorder}) the standard deviation of the power in ring 1 is $\sigma^{ring-1}_{power}=8.2$ nW (13.2 nW) in the \emph{strong dimerization} case and $10.5$\,nW (20.4 nW) \emph{weak dimerization} case. Since a strong signature of topological protection on the SSH model is the localization of light in one of the sublattices, we can also quantify the variation of the power in the even rings in the presence of disorder, which remains very close to zero in the \emph{strong dimerization} case ($\sigma^{even-rings}_{power}=$ 1.3\,nW  and 2.4\,nW for low and high disorder respectively) and it becomes slightly larger in the \emph{weak dimerization} case ($\sigma^{even-rings}_{power}=$ 2.1\,nW  and 4.7\,nW for low and high disorder respectively).

As opposed to conventional topological photonic platforms in which a proper robustness study would require the fabrication and measurement of a large number of devices, this platform allows for accurate quantification of the robustness against disorder in the coupling on the same chip by just software reprogramming. 

\section{2D Topological Photonics in the Programmable Mesh}
\begin{figure}[ht!]
\centering\includegraphics[width=\linewidth]{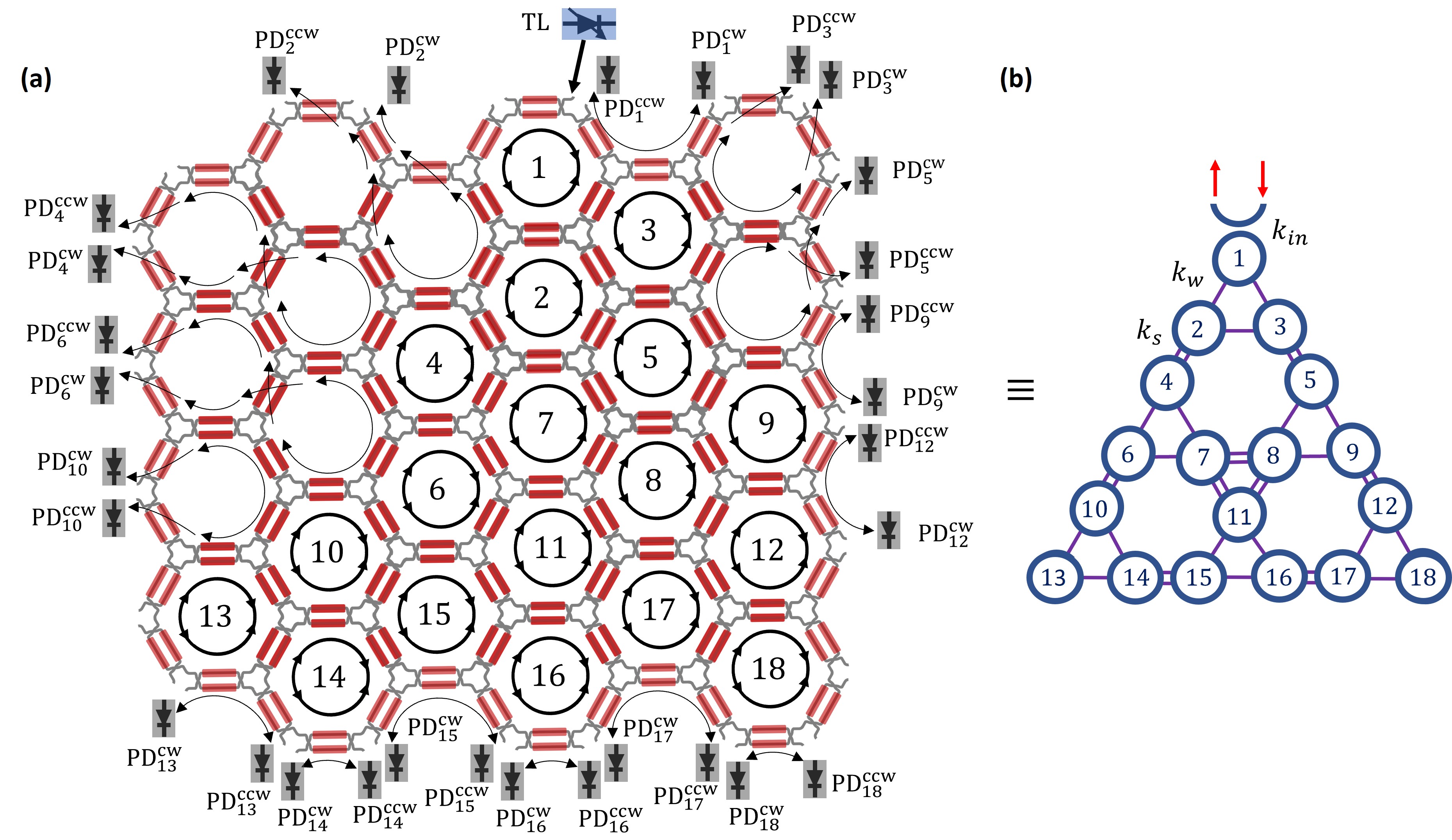}
\caption{(a) Schematic of the programmable mesh on the simulator and reconfiguration for 18 coupled ring resonators, TL: off-chip tunable laser, PD\textsuperscript{(c)cw}: clockwise and counter-clockwise off-chip photodetectors, (b) Implemented 2D breathing Kagome lattice. }
\label{fig:sim_mesh}
\end{figure}

To illustrate the versatility of programmable integrated platforms in the context of topological photonics, we now implement  higher-order topological insulator (HOTI) based on a breathing kagome lattice. The kagome lattice is a 2D model consisting of corner sharing triangles with opposite orientations. While the tight-binding model of the kagome lattice exhibits graphene-like Dirac bands, a band gap opens when the coupling strengths between the sites in different triangles alternate. This is known as the breathing kagome lattice which has been shown to host higher-order topological corner states in a variety of settings \cite{Ezawa2018, peterson2018quantized,serra2018observation,imhof2018topolectrical}, including photonics \cite{xie2019visualization, Chen2019corner, ota2019photonic, mittal2019photonic,elHassan2019corner}. Here, we implement a fully reprogrammable breathing 
Kagome lattice by reconfiguring the silicon photonics mesh into a 2D array of coupled ring resonators arranged in corner sharing triangles with the upward pointing triangles and the downward pointing triangles having different coupling strengths, as depicted in Fig.\ref{fig:sim_mesh} (b).  The implementation of such 2D lattice requires 72 PUCs, exactly the number of PUCs available in the silicon photonics chip of our experiments, see \ref{fig:hw_mesh}. However, the rectangular shape of this specific chip prevents the implementation of the model in Fig. \ref{fig:sim_mesh} (b) directly on the hardware, and thus we have implemented this model on a realistic simulator of the mesh \cite{Perez-Lopez2022}. Note that the scalability required for this demonstration is perfectly within the possibilities of the current technology. 

\begin{figure}[ht!]
\centering\includegraphics[width=\linewidth]{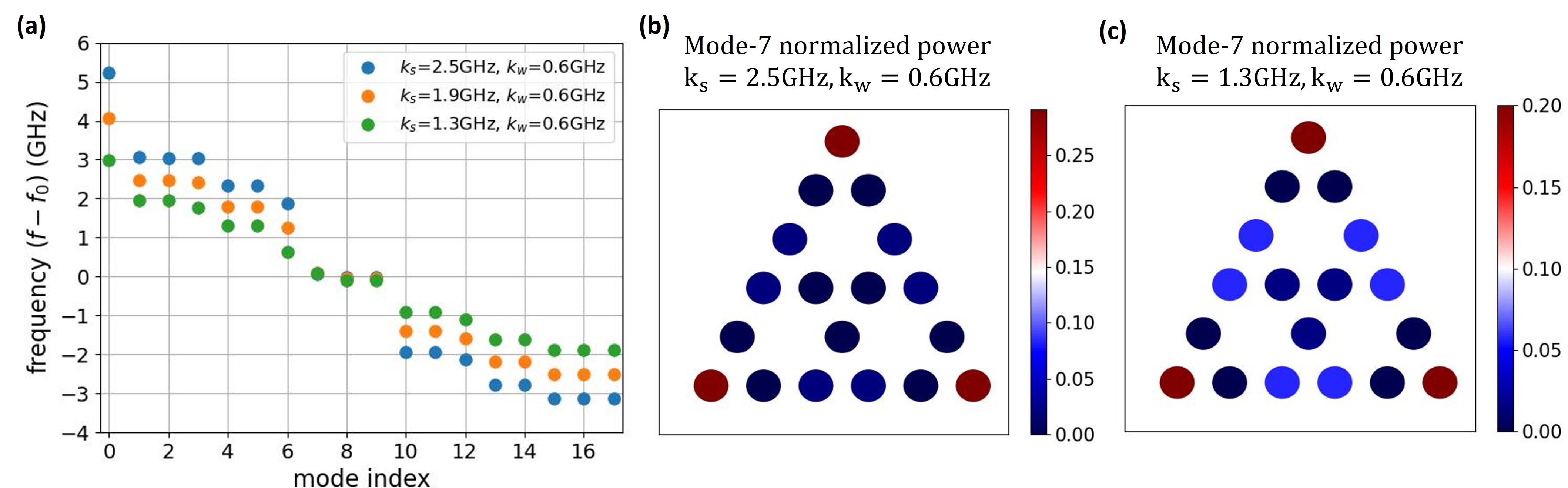}
\caption{(a) Energy spectra of the breathing kagome lattice for three different dimerization patterns: (blue) $k_s=2.5$\,GHz and $k_w=0.6$\,GHz, (yellow) $k_s=1.9$\,GHz and $k_w=0.6$\,GHz, and (green) $k_s=1.3$\,GHz and $k_w=0.6$\,GHz; (b) Normalized power distribution of one of the quasi-degenerate corner states (mode-7) for $k_s=2.5$\,GHz, $k_w=0.6$\,GHz and (c) for $k_s=1.3$\,GHz, $k_w=0.6$\,GHz}
\label{fig:2d_eigen}
\end{figure}

The tight-binding Hamiltonian describing the breathing kagome lattice is
\begin{equation}
H=k_w\sum_{\langle n,m\rangle\in\triangle} a_n^\dag a_{n+1} + k_s\sum_{\langle n,m\rangle\in\triangledown} a_n^\dag a_{n+1}
\end{equation}
where $\triangle$ and $\triangledown$ represent the sites in the upward and downward pointing triangles. The theoretical energy spectra for three different dimerization patterns are shown in Fig. \ref{fig:2d_eigen} (a). We observe three quasi-degenerate energies at $f-f_0\approx0$ that correspond to the energies of the corner states. The power distribution of one of those eigenmodes is shown in Figs. \ref{fig:2d_eigen} (b) and (c) for the stronger and weaker dimerization cases, respectively. It is evident that stronger dimerization leads to stronger light localization at the corners of the lattice.

\begin{figure}[ht!]
\centering\includegraphics[width=\linewidth]
{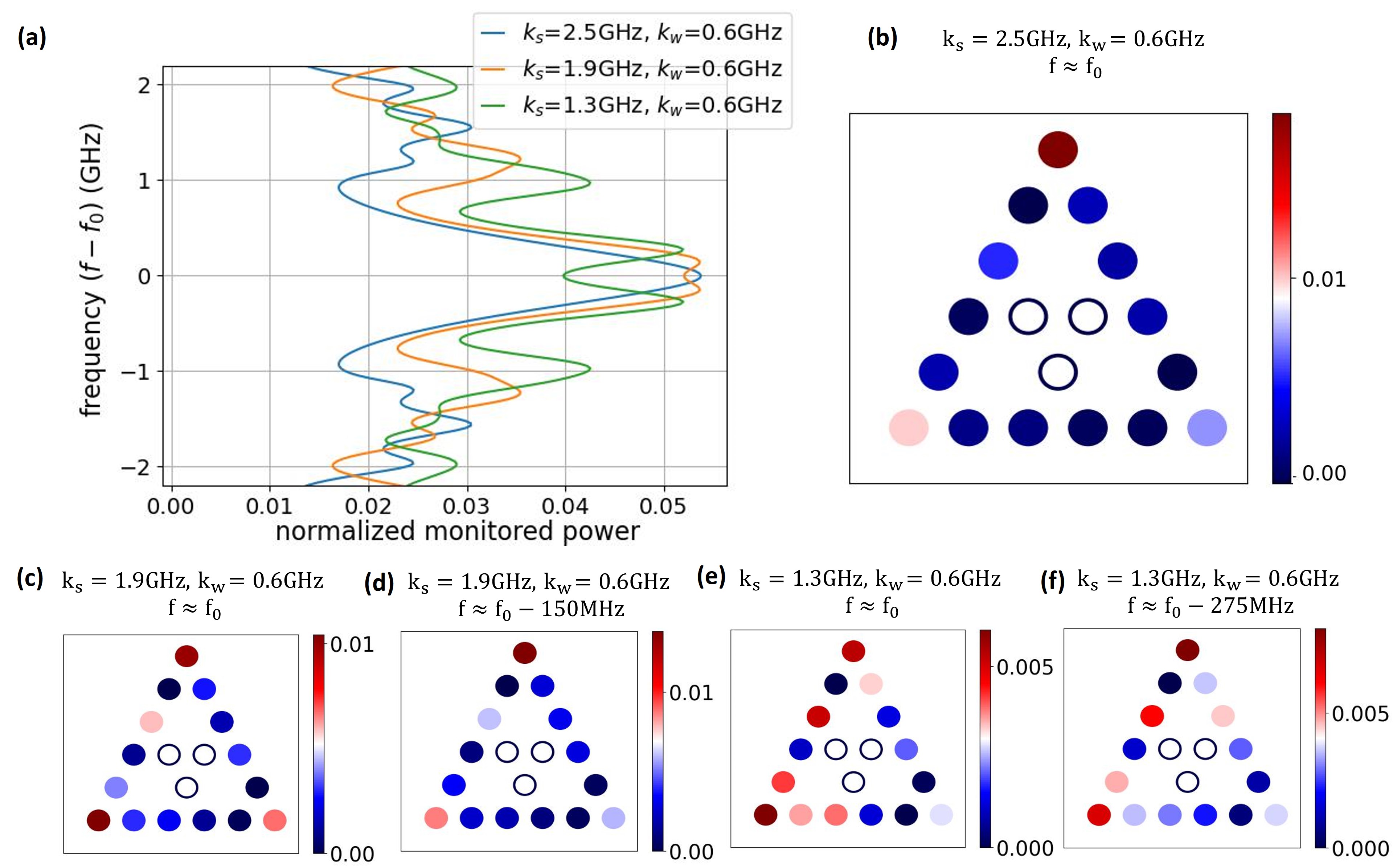}
\caption{(a) Simulated spectrum with various dimerizations, (b-f) Simulated power distribution on the Kagome lattice wit various dimerizations and frequencies.}
\label{fig:2d_result}
\end{figure}

Next, we  simulate the insertion of light in ring 1 and monitor the power in each ring at the edge of the lattice. In the 2D Kagome lattice, light propagates clockwise and counter-clockwise directions inside the resonator, unlike the 1D SSH implemented on hardware mesh. Therefore, each  resonator requires two monitoring ports and external detectors, as shown in Fig. \ref{fig:sim_mesh} (a). First, we vary the input frequency of the laser within a range of $\pm2$ GHz around $f_0$ and sum the monitored power in all the edge rings for each frequency, as shown in Fig. \ref{fig:2d_result} (a). In the case with stronger dimerization (blue line) in Fig.\ref{fig:2d_result} (a) we observe a well defined peak at $f_0$, which indicates that most of the input light populates the corner states and that these states are strongly degenerate. This is confirmed by the power distribution over the edge rings shown in \ref{fig:2d_result} (b), in which the power is strongly localized in the three corner rings. Note that we do not have access to monitoring the bulk rings (rings 7,8 and 11) for the current programmable mesh architecture. However, it is possible to implement monitoring inside the mesh by non-invasive, contactless integrated light probes \cite{clipp}.

Another interesting physical effect occurring in HOTIs under certain conditions is that of light \emph{fractionalization} between the higher-order states \cite{elHassan2019corner}. Given the frequency degeneracy of the three corner states, inputting light in one of the corners is equivalent to exciting an equal superposition of the three corner eigenstates. We can observe some \emph{fractionalization} of light to all three corners, in Fig. \ref{fig:2d_result} (b), the power in all three corners is not exactly equal. We have verified that this is due to the path-related phase differences experienced by the light reaching the bottom left and bottom right corners because of the slightly asymmetric implementation of the lattice in the silicon photonics mesh. This can be remediated by implementing a symmetric mesh (see Supplemental Document Section 2.). 

Subsequently, we focus our attention on the cases with moderate (yellow line) and weak (green line) dimerization in Fig.\ref{fig:2d_result} (a). As the dimerization becomes weaker, so does the degeneracy of the corner states around $f-f_0\approx0$ and this translates into the two sub-peaks observed on the power spectrum monitored on the edge rings. This becomes more pronounced for the weakest dimerization case (green line). Therefore, as illustrated in Figs.\ref{fig:2d_result}(c-g), when the input light has a frequency of $f_0$ the localization on the corner rings is not as strong as at the frequency of the subpeaks ($f_0-0.15GHz$ and  $f_0-0.275GHz$ for the moderate and weak dimerization cases, respectively). For a quantifiable comparison, the percentage of light in the corner rings \rem{compared to the unit input power goes from 1.4$\%$ to 1.9$\%$}\add{increases 36$\%$} when moving from $f_0$ to $f_0-0.275GHz$ in the weakest dimerization case.

\section{Conclusion}
We have proposed and demonstrated that programmable integrated photonics can be used to implement different topological photonics models and to fully reconfigure the behavior of topological modes. In the same platform we have implemented a 1D SSH chain and a 2D HOTI and we have shown full control over the localization and robustness of the edge and corner modes.

The possibility of engineering, not only the coupling rate between sites, but also the phase of such couplings renders this platform readily available for the implementation of a wide variety of topological models, including magnetic-like Hamiltonians that have shown potential in lasers \cite{Bandres2018} and quantum optics functionality \cite{Mittal2018,Mittal2021}. Moreover, the loss of each ring can also be individually and accurately controlled, opening a plethora of possibilities for non-Hermitian topological photonics investigations and devices\cite{guo2009observation,ruter2010observation,ozdemir2019parity,Nasari:23}.

Another enticing future research avenue on this kind of programmable integrated platform is the exploration of lattices with explicitly broken time reversal symmetry (\emph{T}) by time-harmonic modulation of the coupling strength between resonators \cite{fang2012realizing}. A crucial requirement here is that the strength of the modulation must be larger than the decay rate, which translates into the need for fast modulation and low loss technologies. While the current hardware uses heaters to control the coupling and the loss is relatively high, it is within the scalability scope of this technology to introduce high-speed electro-optics phase-shifters and significantly reduce the loss of each cell. This would open the door to the study of a variety of truly non-reciprocal systems at optical frequencies with important fundamental and practical implications.

By showing that a general purpose programmable integrated photonics platform can be used to implement nearly any topological photonics model we hope to accelerate progress in the field, bypassing lengthy design and fabrication cycles and offering a fully reconfigurable platform in which the topological modes are easily tailored and the effects of disorder can be accurately quantified.




\def\url#1{}
\bibliographystyle{IEEEtran}
\bibliography{refs}

\end{document}


\maketitle

\section{Experimental Setup Details and Programmable Mesh Calibration}

All the hexagonal resonators programmed in the mesh consist of six ideally equal PUCs and should, therefore, have equal resonant frequencies. However, fabrication variations on the silicon waveguide cause phase errors \cite{641530}. Additionally, local temperature variations on the processor chip shift the resonance wavelength of the individual resonators \cite{6691908}. Even though iPronics's SmartLight Processor utilizes a thermoelectric cooler packaged with the processor, we observed that resonance wavelengths of the resonators vary, as shown in Fig.\ref{fig:hw_cal} (a). By using two thermoptical phase shifters in the 2$\times$2 programmable unit cell (PUC), (equation (1-3) in the main text), we can tune the resonance wavelength of the individual resonators without disturbing power coupling ratios. We run a calibration procedure before the 1D SSH model measurements. First, we configure the programmable mesh so that the resonators can be excited and monitored one by one. Then, one of the dedicated PUC in each hexagonal resonator is tuned adaptively to match the resonance wavelength of every hexagonal resonator as shown in Fig.\ref{fig:hw_cal} (b).

\begin{figure}[ht!]
\centering\includegraphics[width=\linewidth]{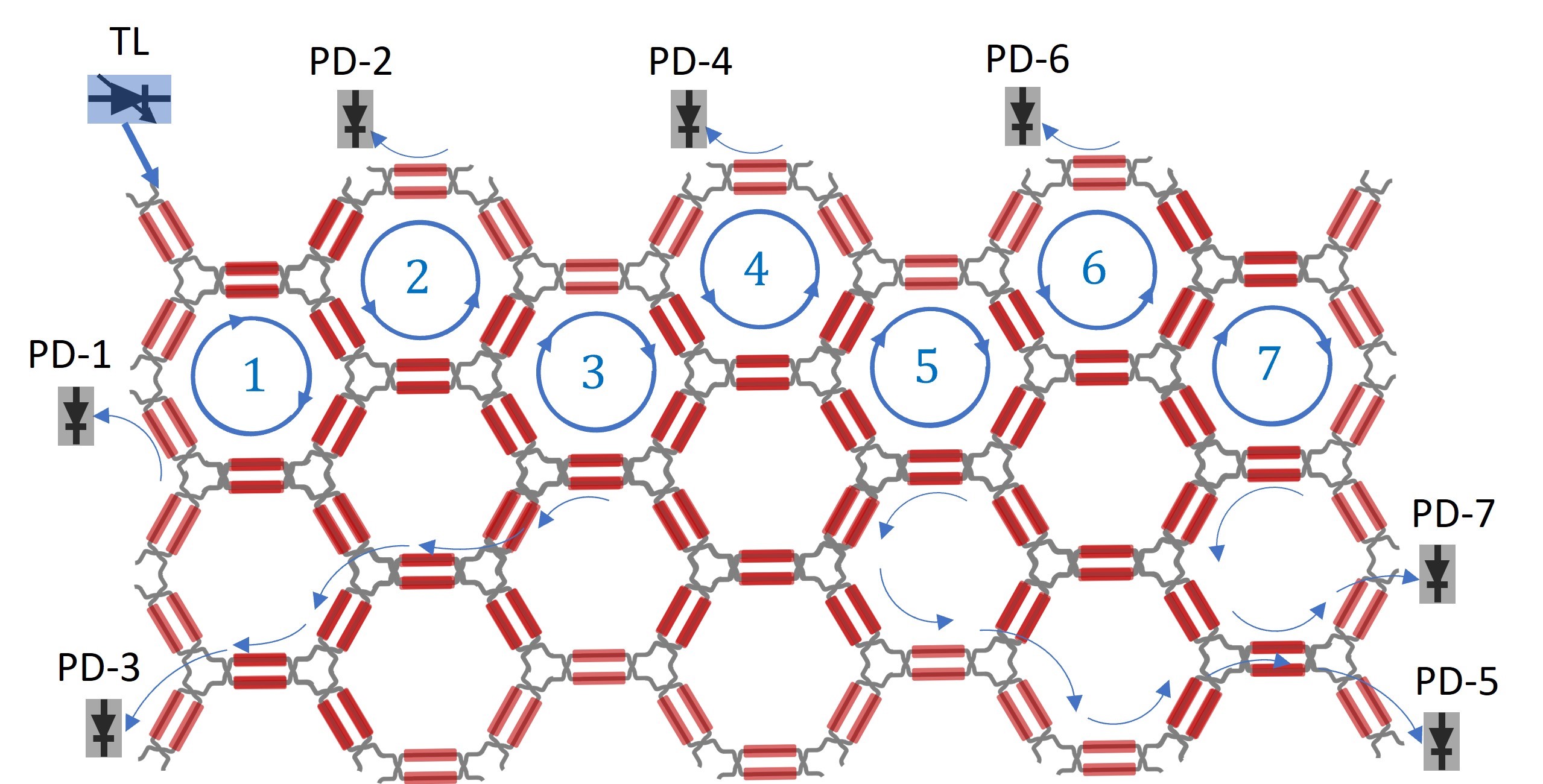}
\caption{Schematic of the programmable mesh on iPronics's SmartLight Processor and reconfiguration for 7 coupled ring resonators, TL: off-chip tunable laser, PD: off-chip photodetector}
\label{fig:hw_mesh_sup}
\end{figure}

\begin{figure}[ht!]
\centering\includegraphics[width=\linewidth]{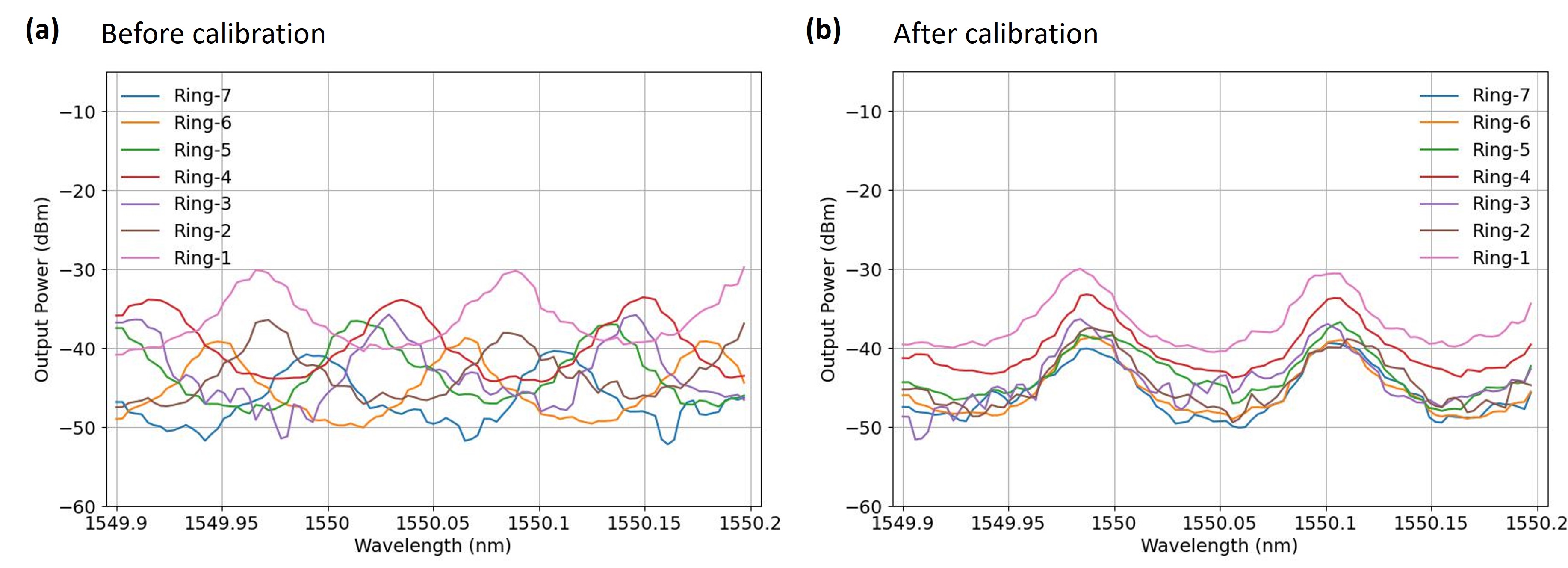}
\caption{Individually monitored ring resonators (a) before and (b) after the resonance wavelength calibration}
\label{fig:hw_cal}
\end{figure}

Fig.\ref{fig:hw_mesh_sup} illustrated the experimental setup, excitation port, and monitoring ports. Because the output ports are located only on the edges of the mesh, the power tapped out of certain resonators (specifically resonators 3, 5, and 7) need to travel through multiple external PUCs to be monitored. Since PUCs have $0.5\pm0.05$\,dB insertion loss in this specific hardware implementation, longer paths to the photodiodes result in non-negligible additional loss. Therefore, we recorded the monitored power at the resonance wavelength for each resonator and subsequently postprocessed the measurements of the rings in the 1D SSH model to compensate for the additional extrinsic loss associated with the longer paths to reach PD-3, PD-5, and PD-7. 

We use a C-band tunable laser (TL) with 0.003\,nm tuning resolution. The TL sweep from 1549.9\,nm to 1550.2\,nm and monitoring photodetectors (PD) measure the optical power values. The sensitivity of the PDs is -70\,dbm. For monitoring purposes, we tapped 1$\%$ of the power by one of the available PUC of the hexagonal resonator. SmartLight Processor is programmed through the Python interface developed by iPronics. Through the interface power coupling ratio ($\sin^2(\Delta)$) of the PUCs and cross-phase values ($\theta$) are set. The power coupling ratio is a value between "0" and "1". "0" represents the bar state for the 2$\times$2  unit, while "1" is the cross state. The coupling ratios between the resonators are set accordingly. We refer to the following formula derived similarly in the supplemental document of the article \cite{Hafezi2013}, 
\begin{equation}
k=\sin^2(\Delta)\frac{FSR}{4}
\end{equation}
to calculate coupling rates presented in the main text. We measured the free-spectral range of the hexagonal resonators as 14.6\,GHz. We set input PUC's power coupling ratio as 0.8 ($k_{in}=2.9GHz$) to inject most of the TL output power into the 1D SSH model.

\section{Simulation Details and Path-Symmetric 2D Kagome Lattice}

\begin{figure}[ht!]
\centering\includegraphics[width=0.9\linewidth]{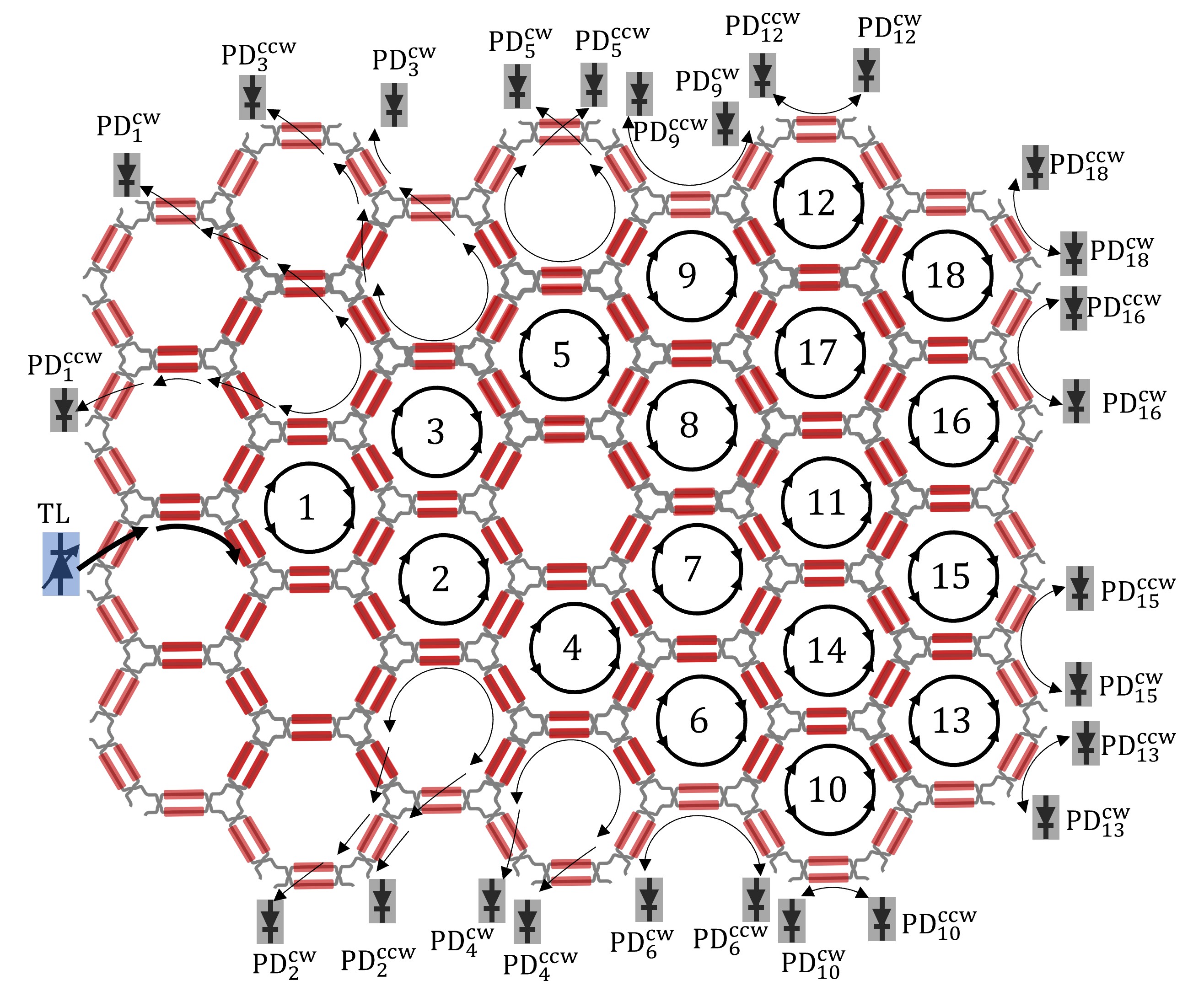}
\caption{Schematic of the simulated path-symmetric 2D Kagome lattice on the programmable mesh}
\label{fig:sym_mesh}
\end{figure}

We observed that various implementations of the 2D breathing Kagome lattice might result in slightly different power localization even though all the simulation settings are equal. Optical path differences between the input PUC and the corner resonators are the main cause of these variations. To verify this, we implemented the symmetric configuration in Fig.\ref{fig:sym_mesh}, which resembles more closely the ideal 2D Kagome lattice in terms of light paths and results in a more equal power distribution on the corner rings. Note that at the time of these studies, the simulator can only achieve a maximum of 5 hexagons in the vertical direction. Therefore, the power on ring-17 and ring-14 in this symmetric configuration could not be monitored, which is the reason for showing the more asymmetric case in the main text. Recent work by Sanchez \textit{et. al} \cite{Perez-Lopez2022} opens a path for arbitrary size programmable mesh simulations in the future.

\begin{figure}[t!]
\centering\includegraphics[width=0.9\linewidth]{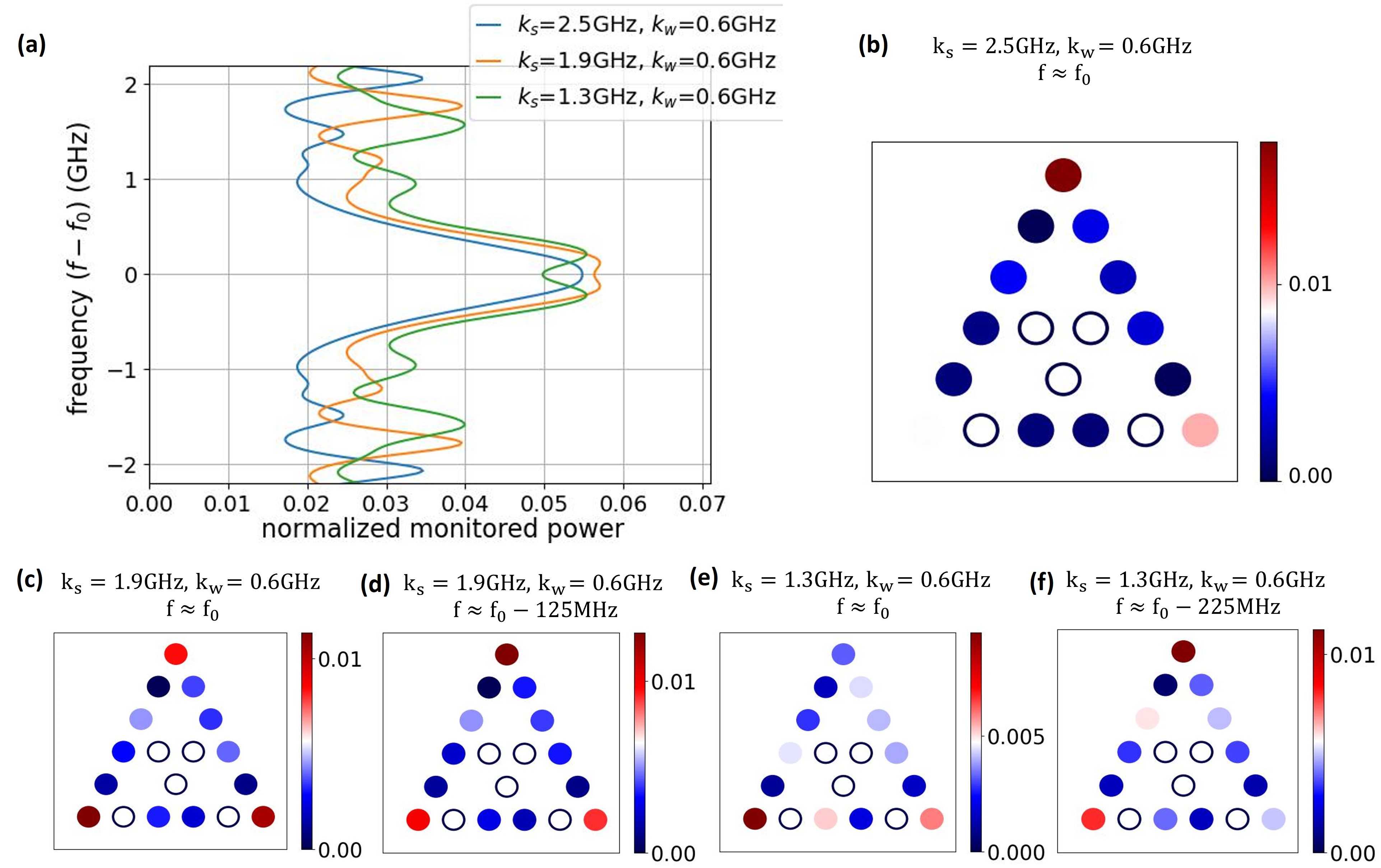}
\caption{(a) Simulated spectrum with various dimerizations, (b-f) Simulated power distribution on the Kagome lattice wit various dimerizations and frequencies.}
\label{fig:sym_mesh_result}
\end{figure}

Only the 2D Kagome lattice resonators' PUCs in the simulator have 0.1\,dB insertion loss, while the routing PUCs for monitoring paths are lossless. The tuning precision of the laser is 0.2pm ($\sim25MHz$) around 1550\,nm. The FSR of the simulated resonators is 12.5\,GHz. The coupling rate for the input PUC to the system is set to $k_{in}=1.6$\,GHz. Fig.\ref{fig:sym_mesh_result} presents the simulation results for the path-symmetric lattice. While we observed similar behavior of "degeneracy" in the spectrum, corner rings 13 and 18 have more localized power than the main text Fig.6.

\clearpage
\def\url#1{}
\bibliographystyle{IEEEtran}
\bibliography{supp}
